# Streamline Intelligent Crowd Monitoring with IoT Cloud Computing Middleware

Alexandros Gazis [1,*] and Eleftheria Katsiri [1,2]

[1] Department of Electrical and Computer Engineering, School of Engineering, Democritus University of Thrace, 67100 Xanthi, Greece; ekatsiri@ee.duth.gr
[2] Institute for the Management of Information Systems, Athena Research & Innovation Center in Information Communication & Knowledge Technologies, 15125 Marousi, Greece
* Correspondence: agazis@ee.duth.gr

**Abstract:** This article introduces a novel middleware that utilizes cost-effective, low-power computing devices like Raspberry Pi to analyze data from wireless sensor networks (WSNs). It is designed for indoor settings like historical buildings and museums, tracking visitors and identifying points of interest. It serves as an evacuation aid by monitoring occupancy and gauging the popularity of specific areas, subjects, or art exhibitions. The middleware employs a basic form of the MapReduce algorithm to gather WSN data and distribute it across available computer nodes. Data collected by RFID sensors on visitor badges is stored on mini-computers placed in exhibition rooms and then transmitted to a remote database after a preset time frame. Utilizing MapReduce for data analysis and a leader election algorithm for fault tolerance, this middleware showcases its viability through metrics, demonstrating applications like swift prototyping and accurate validation of findings. Despite using simpler hardware, its performance matches resource-intensive methods involving audio-visual and AI techniques. This design's innovation lies in its fault-tolerant, distributed setup using budget-friendly, low-power devices rather than resource-heavy hardware or methods. Successfully tested at a historical building in Greece (M. Hatzidakis' residence), it is tailored for indoor spaces. This paper compares its algorithmic application layer with other implementations, highlighting its technical strengths and advantages. Particularly relevant in the wake of the COVID-19 pandemic and general monitoring middleware for indoor locations, this middleware holds promise in tracking visitor counts and overall building occupancy.

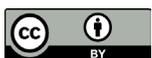

## 1. Introduction

In recent days, the forthcoming Fourth Industrial Revolution, particularly the widespread use of artificial intelligence (AI)—whether in its narrow form (as a means to streamline simple, repetitive processes) or its general form—is poised to alter our daily lives significantly [1,2]. As such, our goal should be to create new systems that will enable society to flourish, guiding this digital transition through sophisticated computer architectures and infrastructures to influence our everyday decisions. This shift is driven by the escalating volume of data produced by an increasingly interconnected network of devices, commonly known as the Internet of Things (IoT). The IoT constitutes a network where various machines are interconnected, allowing them to interact and process data collected from numerous devices, typically sensor nodes. The emerging trend, the Internet of Everything (IoE) [3], expands the capabilities of the IoT, providing the opportunity not just for device and machine connectivity, but also for integrating people and overall processes into the data circulating within the existing network [4]. Despite the uncertainties of the future, the pivotal concepts remain the IoT and the Machine to Machine (M2M) communication protocols, alongside the growing utilization of data points from various nodes, whether referring to complex system infrastructures or embedded, low-cost, and low-power sensory devices [5].

Currently, the focus is on IoT and M2M communication, exploring new methods to leverage the increasing volume of data and big data being generated. In this context, terms like IoT, big data, and cloud computing are frequently mentioned by scholars to emphasize the need to process and integrate data from multiple sources into a unified, vast network of devices. This necessitates the development of novel M2M communication methods to expedite the extraction, loading, and transformation of large volumes of data, whether originating from machine endpoints or human–machine interactions [6]. Recent studies highlight one of the era's challenges: the computing resources required to extract and process this data volume [7,8]. Whether through AI methods or new tools utilizing big data for accuracy or predictions, a critical consideration is the overall operational and execution budget [9,10]. Moreover, since these systems often rely on existing architectures, they are generally monolithic applications, not favoring decentralized solutions or leveraging concepts [10], like edge computing or ubiquitous/pervasive computing [11].

The term "pervasive computing", coined in the 1990s, signifies the shift from monolithic infrastructures to "ambient intelligence", or "everywhere computing". This aims to equip devices with the capability to be smart through an abstract layer or taxonomy of processing technologies that do not overlap. For instance, new applications employing ubiquitous computing take advantage of internet support to propose advanced middleware applications that provide a structured approach for operating systems, sensors, microsensors embedded in devices, and I/O processes to connect and exchange information [12–14]. Since the 1990s, the relevance of this term has only grown, in line with Moore's Law [15,16], which suggests that as years pass, the cost of devices decreases while CPU capabilities increase rapidly. Additionally, Huang's Law [17] observes that advancements in graphics processing units (GPUs) also follow a similar pattern, offering a chance for device properties to evolve at a steady and even faster rate than CPUs. These observations suggest that existing low-cost and low-power devices could offer alternative means to host and develop more complex and resource-intensive operations [18]. Notably, these capabilities are utilized by "edge computing", an emerging technology that enables networks and devices to process data closer to where it is generated, either embedded on the node or after real-time processing on the computer's node [19].

Ubiquitous computing suggests that systems must transition from monolithic infrastructures towards more system-oriented approaches, embracing principles such as "don't repeat yourself" and SOLID class design principles [20]: Single Responsibility, Open–Closed, Liskov Substitution, Interface Segregation, and Dependency Inversion [21]. This involves developing a middleware that leans more towards an object-oriented rather than a procedural or transactional middleware architecture, or attempting a message-oriented infrastructure to ensure that each object entity performs a unique job but remains flexible for extension without modification. Following the Liskov principle [22], it should be substitutable for its base or parent object, aiding in segregating concerns and reducing dependencies on other processes and methods. Consequently, this fosters the creation of more abstract entities, or more precisely, general abstraction software layers, often referred to as middleware, which do not rely on low-level modules but instead provide a general, multipurpose interaction layer [23].

The term is closely associated with middleware, a concept introduced in the early 1980s to define a software pipeline—an operation, a process, or an application situated between the operating system and the end user [24]. This concept was developed to address the need to modernize legacy systems through the separation of concerns and the expansion of existing layers, facilitating generic communication among various applications and software components within distributed, interconnected networks. Drawing from this, ubiquitous computing introduces the following layers as a means to form a new abstraction of a monolithic system [25,26]:

- Environment layer: to monitor the resources and capabilities of computing systems, suggesting extensions or modifications of existing properties and capabilities.
- Environment management layer: to oversee all relevant resources and enhance the reliability and continuity of operations.
- Task management layer: to manage the tasks, content, and requirements of services and dependencies of objects.

In our article, we aim to expand upon these layers of abstraction to propose a novel middleware for use in wireless sensor networks (WSNs) that rely on low-power and low-cost (mini) computer devices. Our article is structured as follows: we begin by stating the aims and objectives of our study. Then, we present a literature review of WSNs and their extensive application across various industry sectors. Next, we explore the origins of MapReduce and present a case study of a historically significant building in Greece. We then describe our experimental setup selected to test our case study and perform our algorithmic processes, including a fault-tolerant MapReduce operation for WSNs. This algorithm, which is detailed thoroughly, serves as the foundation of our middleware implementation for WSNs. Finally, we discuss the rationale behind our middleware, developed through the use of low-power and low-cost devices such as Raspberry Pis, presenting our results and benchmarks for the proposed system and suggesting avenues for future research.

## 2. Aims, Objectives, and Technical Contribution

The aim of this article is not to act as a review paper on the subject but rather to present a cost-effective and scalable IoT middleware system for crowd monitoring. The development effort provides a top-down approach to all the layers and the techniques used to simplify and optimize the workflow of the software entities. The primary objectives include optimizing the development steps, defining the system architecture, and detailing the processing of experimental data. Also, we provide a comprehensive case study, complete with actual images and building blueprints, to illustrate the system's application. The structure of each middleware layer, with a focus on crowd monitoring, is also defined, emphasizing using low-power and low-cost solutions in terms of computing resources. Furthermore, the implementation of MapReduce is discussed for its role in efficient data processing and output assessment as evident from the benchmarks and capabilities of the proposed IoT middleware that is evaluated.

The technical contribution of this work is highlighted by the ability of each mini-computer and sensor node to function interchangeably as either a client or a server. This versatility, combined with MapReduce implementation, allows for the cost-effective and reliable integration of data from multiple endpoints within the IoT network. In contrast to conventional methods that mainly rely on AI or audiovisual analysis (image/video objects)—often requiring significant computing resources for operation and processing—this approach facilitates rapid prototyping and scalable solutions across various endpoints (mini-computers). Lastly, providing an algorithm for operation, a type of fairness index for connectivity status is introduced to prevent single points of failure, ensuring continuous monitoring of availability and downtime, similar to mainframe computing transaction middleware. As such, by fusing multiple technologies, including MapReduce, wireless sensor networks (WSNs), and crowd sensing applications, the proposed middleware aims to suggest optimized processes that minimize overall computing costs and ensure effective crowd monitoring.

## 3. Related Work

The initial step in designing our middleware application was to understand and define what constitutes a wireless sensor network (WSN) and its characteristics [27]. In recent years, following the exponential growth of smart sensory devices, the term WSN and the emergence of such sensor networks have seen a gradual increase. WSNs comprise numerous small or microsensory devices that enable the monitoring of both physical and environmental factors [28]. These sensors are cost-effective to purchase, calibrate, and operate, and they are also energy-efficient [29]. Given their small size, affordability, and intelligence, WSNs are often used interchangeably with the IoT. This term describes a vast network of entities (devices, sensors, nodes, edges) operating within the same network and incorporating cloud computing and big data to facilitate the gathering, extraction, and loading of information among these devices [30]. The advancements following Moore's and Huang's Laws [17] have enhanced the computational power of low-cost computers, while the IoT has provided an opportunity to integrate data from multiple low-maintenance sensors into a large, interconnected network that processes data. Thus, WSNs go beyond mere Machine-to-Machine communication; they aim to not only exchange large volumes of data at high speed and in real time but also provide a global network for connectivity, data analysis, and interaction. Connectivity alone is not achieved through the IoT but rather the secure network of microsensors and devices that interact and share data [31].

Wireless sensor nodes, the edges of WSNs, are defined as compact devices comprising a processor, storage, transceiver, power supply unit, actuator, and general sensor devices for calibration and data signal processing. These nodes, like WSNs, are low-cost and fall into categories such as biochips, nanosensors, and micro-electromechanical systems (MEMSs) [32,33]. Biochips process multiple biochemical reactions on a single chip; nanosensors detect and interact with physical or chemical stimuli at the nanoscale; MEMS monitor and control mechanical systems at the microscale. WSNs employ sensors from these categories to provide data on temperature, humidity, motion, vibration, pressure, light, radiation, etc. A WSN's primary function is to have a sensor node that can originate data and act as a data router, monitoring applications via sensor devices embedded in the node while sending data to its core component (sink) to track and monitor event changes [34].

Moreover, middleware, a term often associated with WSN solutions and architecture, describes decentralized solutions for a set of different but not independent systems. Initially coined in the 1980s, it signified a departure from monolithic applications, like early computer systems, which bundled multiple software entities into one large application [24]. Early e-commerce platforms, for instance, lacked scalability and agility due to their large architectural designs. These systems typically employ a transactional software approach, requiring the processing of multiple synchronous processes for optimal operation. Middleware first emerged as a cluster capable of processing multiple synchronous/asynchronous transactions in a distributed manner, introduced primarily in mainframe computing and by significant financial institutions for transactions and credit card payments. Middleware architecture evolved to suggest message-oriented systems, such as message queues or message parsing architectures, which, beyond synchronous/asynchronous communication, introduced a publish–subscribe pattern with an intermediary broker to manage the communication of devices and edges. This architecture is widely used as it provides a top-down approach of a leader–follower or a decentralized architecture that facilitates interactions within components with agents and actors used to orchestrate the flow of data and final processes in ETL. In WSNs, this application is suggested to improve latency and reliability at the node while considering bandwidth limitations and the overall system's productivity, scalability, mobility, and control. It also provides systems and services with reliable runtime operations, safer code execution, and a decrease in possible errors and handling as they are application-specific entities, and thus easier to expand, maintain, and comprehend. Other middleware architectures consist of remote and local architectures available only to connect, pass, and receive requests and responses of asynchronous operations (procedural computing) for predefined services and object-oriented middleware. This last category is notable as it adopts a more general approach to software entities' responsibilities and capabilities, focusing more on the attributes, properties, and behaviors of each system. These middleware architectures are also used in WSNs as they embody the principles of object orientation, including abstractions, encapsulation, polymorphism, and inheritance, though they are not typically associated with the outcome of WSNs nor the sensor devices' processes and operations [35]. Additionally, a taxonomy of programming models for WSN middleware is presented, highlighting the importance of power and resources, scalability, mobility, dynamic network topology and organization, real-world integration and implementation, and broader considerations like data aggregation, quality of service, and security. In the coming years, middleware architectures will be developed focusing on case-specific applications such as IIoT, Industry 4.0 automation processes, Virtual Reality, embedded system sensor devices, computer protocols and networks, energy optimization, device heterogeneity, the Internet of nano-Things, WSAN, edge computing, and deep learning on microcontrollers (TinyML) [36–38].

Recent research indicates that WSNs typically require a layer of abstraction, i.e., a middleware architecture, to perform and coordinate their various operations, whether that is data generation, data mining, and analysis, or even data transmission to other computing devices or sensors. These middleware aim to provide a standard means of interaction and operation to ensure reliability and scalability, and to provide monitoring and fault tolerance during operation. WSN middleware exists in many aspects of computer science, including artificial intelligence, computer systems and networks, security, database systems, human–computer interaction, vision and graphics, numerical analysis, programming languages, software engineering, bioinformatics, and the theory of computing [39]. WSNs are also employed in applications related to urban [40] and rural [41] projects, outdoor [42] and indoor [34,43] tracking, environmental monitoring [44], disaster monitoring [45], water quality [46], habitat monitoring [47], traffic monitoring [48], earthquake detection [49], volcano eruption [50], agriculture [51], weather forecasting [52], smart water management [53], smart cities [54], smart health [55], smart energy management [56], smart traffic management [57], smart retail [58], smart security [59], smart waste management [60], smart education [61], and smart homes [62].

## 4. Materials and Methods

### 4.1. MapReduce

With the exponential increase in data volumes in recent years, both scientific computing communities and enterprises have encountered the challenge of processing vast quantities of data efficiently [63]. In response to the growing need to manage these large data volumes, the MapReduce programming model has been introduced, offering a high-level abstraction tailored for data-intensive computing tasks. Concurrently, cloud computing has emerged to provide seamless access to extensive computing resources, networking, and storage capabilities, ensuring that applications can effectively handle large datasets. Owing to its wide range of applications and benefits, the MapReduce framework has been applied across various domains [64,65].

Developed by Google, MapReduce is recognized as a pivotal model for big-data analytics, capable of managing complex systems that typically involve massive datasets or multiple points of data aggregation and real-time big-data flows [66]. It is a data-parallel programming model designed for the processing, generation, and analysis of distributed computations over large datasets, executing these tasks on clusters of commodity servers [67,68]. Although initially developed by Google, its principles of parallel and distributed processing, in conjunction with Apache's Hadoop, have become the standard software systems for big-data applications due to their open-source implementation. The primary goal of MapReduce was to simplify data parallelization, load balancing, and distribution through an easily accessible library [66]. It is noted for its flexibility, simplicity, scalability, and excellent fault tolerance, as it requires only the tasks on failed nodes to be restarted, although it can also be costly [69].

MapReduce's applications, leveraging its features and benefits [70], span data mining and extraction for reports [71], big-data graphical computation [72], machine learning challenges [73], statistical machine translation [74], spam detection [75] satellite image data processing [76], and problem clustering [77], among others. MapReduce operates through a combination of map and reduce functions, which together handle machine failures, parallelize computations across vast clusters, and facilitate inter-machine communication scheduling [78]. Users write both maps and reduce functions to process input key/value pairs and generate output key/value pairs. The map function creates intermediate key/value pairs, while the reduce function merges all intermediate values associated with the same intermediate key. The MapReduce library assists the reduce function in merging by supplying intermediate input values via an iterator, enabling the handling of large value lists that exceed memory capacity and are stored in the cloud [79].

The architecture of MapReduce encompasses three main phases: mapper, reducer, and shuffle. The leader node assigns input data to the mapper, which processes the data to produce intermediate key/value pairs. The reducer then combines these intermediate keys and values for a particular key into a smaller set of values. The shuffle phase involves transferring large volumes of data from all map nodes to reduce nodes, executed by the shuffler, which moves data from the mapper disk, not the main memory. This step is crucial as sorting intermediate results by their keys simplifies grouping [80]. This parallel model is highly effective for large-scale data analysis using several cluster machines, offering a robust framework for handling big-data challenges.

The user-defined map and reduce functions are structured as

- **map (k1, v1) → list (k2, v2)**

  and

- **reduce (k2, list (v2)) → list (v2))**

  Meaning:
  STEP 1: Abstract data as key/value pairs to a map function as follows:

$$(key, value) = (k_A, V_A) = (k_{A1}, V_{A1}) (k_{A2}, V_{A2}) \quad (1)$$

$$(key, value) = (k_B, V_B) = (k_{B1}, V_{B1}) (k_{B2}, V_{B2}) \quad (2)$$

STEP 2: Sort/group the function output as follows:

$$(k_{A1}, V_{A1})(k_{A2}, V_{A2}) \ldots \to k(V_A, V_B) = k_N V_N \quad (3)$$

The term MapReduce encompasses both a programming model and its corresponding runtime environment, designed to efficiently manage the execution of both Map and Reduce functions. This execution mechanism is coordinated by two types of entities [81]:

- JobTracker: Acts as the leader, overseeing the complete execution of the submitted job.
- Multiple TaskTrackers: Function as followers, each executing a portion of the job.

For every job initiated within the system, a single "JobTracker" operates from the "NameNode", while multiple "TaskTrackers" are deployed on "DataNodes" [82]. The MapReduce framework has proven to be a pivotal tool in the processing and management of large data volumes, offering a straightforward and effective approach to big-data challenges, particularly in cloud computing environments. MapReduce's advantages are manifold, including its capability to process vast datasets, support data-intensive computing tasks, and accommodate various programming languages.

*4.2. Case Study: Historical Building in Greece*

The case study of this article focuses on a historical building on the outskirts of Xanthi, Greece, the birthplace and childhood home of Manos Hadjidakis (also spelled Hadzidakis), which holds significant cultural value. Hadjidakis, a distinguished Greek composer and music theorist, is celebrated as one of the most influential figures in both Greek and European music history. Alongside Mikis Theodorakis, Hadjidakis is credited with pioneering "Entekhno" music, a genre that melds orchestral music with elements of Greek folk rhythms and melodies, often setting lyrical themes to poetry [83].

After years of neglect, the Greek government has recently undertaken a comprehensive renovation of this notable building, transforming it into a cultural center for the local community. The building is renowned for its architectural details, including stunning frescoes, ceiling paintings, mosaic floors, and stained glass windows, all featuring a blend of neoclassical and Baroque styles. Spanning three floors and occupying a plot of 1317 sq. m, the structure was initially erected in the late 18th century by Isaac Daniel, a wealthy stockbroker and tobacco trader. Constructed in two phases, it was completed in 1829, situated on a hill offering expansive views of the city. Following Daniel's death, ownership transferred to his children, who, unable to afford the inheritance tax, relinquished the building to the Greek Ministry of Finance, which repurposed it as a tax office. Post Greek Civil War in the 1950s, it served as a city garrison before being converted into the cultural center it is today, as illustrated in Figure 1, which showcases the exterior of the Hadjidakis residence [80].

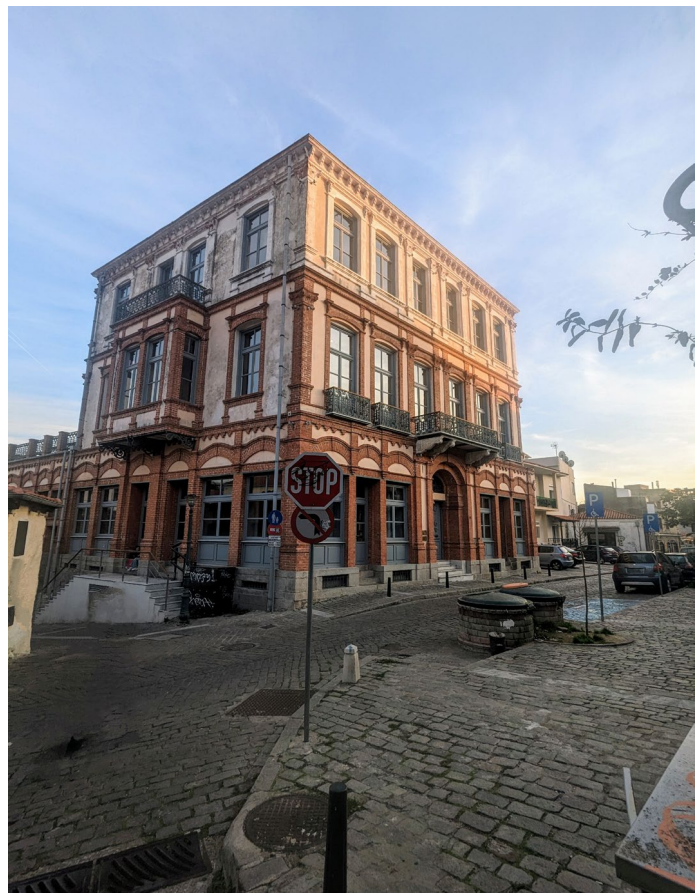

**Figure 1.** Outside view of M. Hatzidakis's residence.

Our case study focuses on the building's main exhibition floor, detailed in Figure 2, where we have installed radio frequency identification (RFID) readers to monitor visitor movement within the first floor, excluding the terrace and restrooms for privacy and relevance reasons. Utilizing the blueprint from Figure 2, we simulated data for hosting art events and exhibitions, leveraging this space's role as a cultural hub for various activities, including exhibitions and concerts. Our system, a middleware solution, integrates with a cloud computing architecture and a wireless sensor network (WSN) IoT infrastructure, ensuring an effective, accurate, scalable, low-cost, and reliable visitor monitoring method.

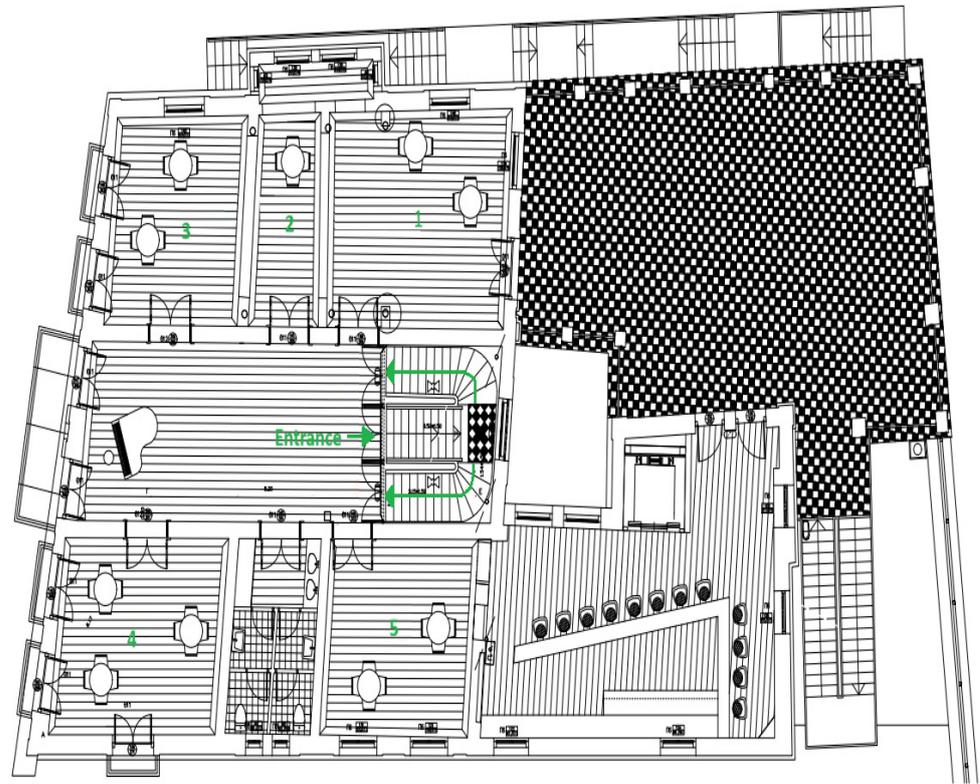

**Figure 2.** Floor plan of Hatzidakis's residence.

*4.3. Case Study: Experimental Setup*

Upon entering the building, visitors are assigned one of three RFID tags indicating "man", "woman", or "other"; the latter option was introduced to accommodate guide dogs for people with disabilities and to consider future scenarios. These RFID tags, encoded upon entrance, allow for the monitoring of visitors in emergencies and, in the wake of the COVID-19 era, to track their movements throughout the floor.

As depicted in Figure 2, each room is outfitted with low-power, cost-effective computers responsible for data acquisition and aggregation. These devices can function as either a server or a client depending on the system's status, connecting to RFID readers and communicating with a central system via socket programming. These units may serve as mappers or reducers—either monitoring RFID activity in their designated room and scanning the area at predetermined intervals (mapping) or calculating the overall visitor count (reducing). Our setup includes two RFID readers at each room's entrance, ensuring detection regardless of the tag's placement on the visitor's arm, using industry-standard, waterproof, and dust-resistant wristband RFID tags.

The operation frequency of our system can reach up to 134 kHz, allowing for a scanning range of 20–30 cm. This setup ensures accurate measurements at doorway entrances. For larger entrances or areas like building entry points, the frequency can extend up to 13.56 MHz to achieve a scanning range of up to 1 m. According to the Centers for Disease Control and Prevention, these solutions are optimal given the average person's arm span (35–41 cm depending on gender) and the typical door width of 1 m [84]. This approach ensures coverage even if individuals do not pass directly through the center of an entrance, prioritizing a low-cost, efficient alternative to more complex and resource-intensive audiovisual systems.

The computing devices employed include Raspberry Pi models 4 and 3B, featuring ARM Cortex-A53 CPU1 with 1 GB and ARM Cortex BCM2711 at 1.5 GHz with 4 GB, respectively. Early tests also utilized Nvidia's Jetson, which offers performance comparable to the Raspberry Pi 4 model [85,86].

*4.4. Fault-Tolerant WSN Algorithm and MapReduce Implementation*

The core of our software system is its application layer, a headless Java implementation that operates continuously. Periodically, it verifies the server's availability and functionality. If the server is found to be non-operational, it initiates a leader election algorithm to designate a new leader, effectively transforming one of the system's worker nodes (clients) into the server role. In the context of our wireless sensor network (WSN) case study, all computers must be network-connected, whether remotely or locally. Upon startup, each computer attempts to connect with the remote database infrastructure, registering its node ID and IP address. This database maintains the network properties (port, IP address) and node IDs for both the leader node (server) and the follower nodes (clients), ensuring each node's uniqueness through an integer identifier.

Should there be any missing information upon connection, each mini-computer is programmed to submit the required data to the database or update any existing records. Subsequently, each server re-establishes a connection to the remote database to compare node IDs. The node with the highest ID number is then elected to serve as the server. Additionally, this process allows for customization; users can specify through a flag parameter in the console command at system initiation which computer should assume the server role, offering flexibility in network configuration and management.

4.4.1. Fault Tolerance Algorithm

The application layer of our software system, pivotal to its functionality, is implemented in Java without a graphical interface. It operates continuously, checking every 20 min if the server is operational. If not, it attempts to assume the leadership role instead of merely acting as a client. This routine, essential for maintaining system integrity, especially in a wireless sensor network (WSN), requires a minimum of two connected computers for system operation. Each computer, upon initial startup and at subsequent 20 min intervals, verifies server availability. If a server is unavailable, the system triggers a leader election to appoint a new server from among the worker nodes.

Upon connecting to the network—be it remotely or locally—each computer registers its node ID and IP address in a remote database. This database catalogs the network properties (IP address, port) and node IDs, ensuring unique identification for the server (leader node) and clients (follower nodes). If a computer finds its data missing or outdated in the database, it updates or inserts the necessary information. Subsequently, the system checks if any node's ID number surpasses others; the highest ID assumes server duties. This protocol can be overridden at system startup through a user-specified flag, allowing manual server selection.

The system mandates a 20 min operation cycle, aligning with the average duration of a guided tour (15–17 min), allowing ample time for data transmission and analysis. Within this period, the algorithm facilitates client–server communication: clients generate random visitor data (man/woman/other) and the corresponding room measurements. These raw sensor data, once aggregated and categorized into key/value pairs, are sorted in ascending order on each client and transmitted to the server via a UDP connection.

The server, upon receiving data from clients, waits for the 20 min cycle to conclude before processing the aggregated information. If fewer than two nodes respond, the system attempts to re-elect a server. Otherwise, it consolidates client data into a comprehensive list, divided into segments corresponding to the number of clients. Each client receives a portion of this list for MapReduce operations, with assignments based on their node ID. The results are then sent back to the server for visitor count analysis. If the data are intact and correct, the server updates the remote database with visitor counts by genre and room. A success message is dispatched to all clients, confirming the operation's outcome.

This algorithm's innovation lies in its resilience, not relying on a single node for operation continuity. Unlike Hadoop's architecture, which employs checkpointing for error recovery, our model is tailored for WSNs, where sensor nodes' limitations (speed, storage, capabilities) make traditional checkpointing impractical. Instead, our approach ensures data integrity through network connectivity checks and leader election, proving effective for small-scale operations and serving as a validation for checkpoint methods in scenarios with minimal data flow or medium-sized operations through virtual or horizontal scaling. Our algorithm prioritizes network connectivity, re-evaluating leader–server assignments based on active nodes, and ensuring no data loss or unprocessed information in any operational scenario. Figure 3 illustrates the client–server interaction within our distributed system, highlighting the algorithm's operational flow and its efficiency in handling WSN measurements. Similarly, the algorithmic steps of the proposed solution are presented in Algorithm 1 using a generic software description of each state as presented in Figure 3.

**Algorithm 1.** Figure 3's algorithmic steps of the proposed solution using a generic software description of each state.

```
# Function: run_wsn_monitoring_system()
def run_wsn_Middleware_system():
# Initialize variables
node_id = register_node()   # Register node and get ID
server_id = get_server_id()   # Get current server ID

while True:
# Check server availability every 20 min
if not is_server_available(server_id):
server_id = elect_leader()   # Trigger leader election

# Generate and transmit sensor data every 20 min
data = generate_sensor_data()
send_data_to_server(server_id, data)

# Receive processing instructions from the server
instructions = receive_instructions_from_server()

# Process data based on instructions (e.g., map-reduce)
processed_data = process_data(instructions)

# Send processed data back to the server
send_data_to_server(server_id, processed_data)

# Receive and display a success message
message = receive_message_from_server()
```

**print(**message**)**

# Wait for the next cycle
time**.**sleep**(**20 × 60**)**　# Sleep for 20 min

# Helper functions
**def** register_node**():**　# ... registers node and returns ID

**def** get_server_id**():**　# ... retrieves current server ID

**def** is_server_available**(**server_id**):**　# ... checks server availability

**def** elect_leader**():**　# ... conducts leader election

**def** generate_sensor_data**():**　# ... generates visitor and room measurements data

**def** send_data_to_server**(**server_id**,** data**):**　# ... sends data to server using UDP

**def** receive_instructions_from_server**():**　# ... waits and receives server instructions

**def** process_data**(**instructions**):**　# ... processes data based on received instructions

**def** receive_message_from_server**():**　# ... waits for and receives a message from the server

## Parallel WSN System Process

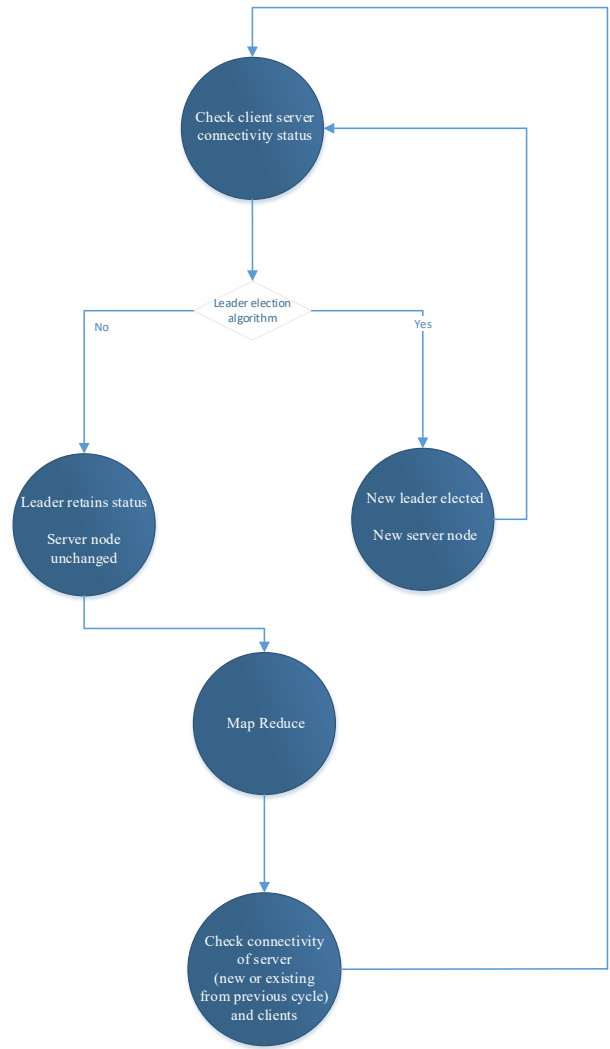

**Figure 3.** Flow diagram of our parallel WSN system processes with MapReduce and leader election.

4.4.2. Map Reduce Visitor Count Rationale

Our MapReduce implementation draws inspiration from the classic word count example, aiming to identify and cluster recurring data within a data flow to count the occurrence of each "word". Initially, data regarding visitors in each room are collected and analyzed to produce an output. The MapReduce operation functions in two modes: visitor counting and room counting. The former categorizes visitors based on their RFID tag identity (man, woman, other—e.g., dog) into key/value pairs, while the latter focuses on counting the total visitors in each room. The first mode is crucial for evacuation strategies by monitoring the number of people within a building, whereas the second mode identifies which exhibition attracts the most attention or, when used together, counts the total number of visitors in scenarios like a pandemic to track virus spread through the movement of individuals carrying the virus and their contact points within specific location rooms.

For our case study, we have chosen to place one computer in each room, as outlined in the presented blueprint, selecting Raspberry Pi models 3 and 4 for their adequate dual and quad-core CPU capabilities and over 1 GB of RAM. Devices are strategically placed in all rooms except the entrance and restrooms to measure passerby activity. An allocated timeframe of 2–5 min within the total 20 min software cycle is provided for the MapReduce operation, facilitating the data aggregation technique. In this setup, server nodes perform mapping, while client nodes execute reduction tasks for each visitor count operation. The rationale behind our MapReduce system is depicted in Figure 4, showcasing the efficient use of Raspberry Pi devices as a proof of concept for both hardware capability and algorithmic functionality.

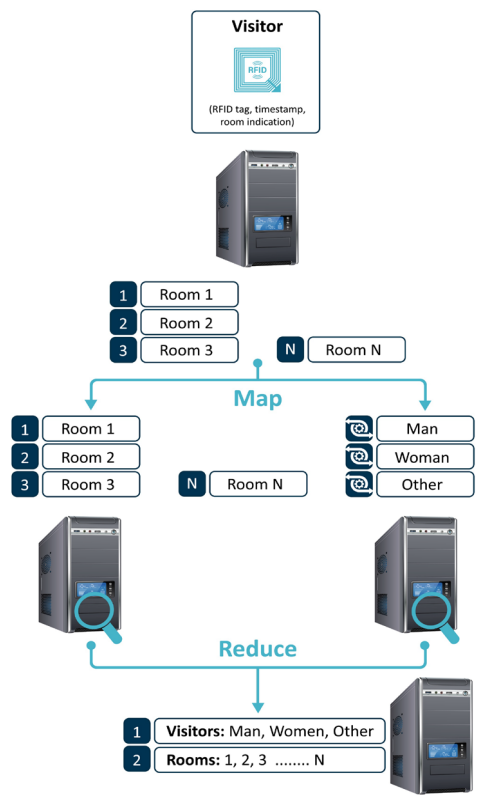

**Figure 4.** MapReduce algorithm rationale for crowd monitoring based on exhibition rooms and RFID tag information.

*4.5. Storage Properties (DB and Information)*

Initially, to store the data collected from sensors, we created .txt files. However, aiming for a more robust and agile data management solution, we considered transitioning to CSV files. Despite their lightweight nature, CSVs lacked a centralized, easily indexable format compared to our daily text file approach. Consequently, we explored more advanced solutions, such as exporting our data into a database (DB) to enable trigger events and complex operations.

Our first strategy involved setting up a local network and hosting the DB on a local computer, utilizing free, open-source software like XAMPP [87]. This software bundle provides a comprehensive framework to start and run a local server (Apache) and interact with our DB through a graphical user interface via PhpMyAdmin.

Upon further consideration, we decided to adopt a cloud-based solution, using db4free—a free cloud-based MySQL testing service [88]. This service allowed us to store up to 200 MB of data, facilitating testing in a production-like environment without the need to establish a local network/server for our database.

Our database comprises two tables: one for the network properties of the operation and another for the final MapReduce results of the visitor count analysis. The first table includes two columns—an integer and a varchar—storing the node ID and network properties of each computer node. The second table captures the actual results of the MapReduce operation, detailing the total count of visitors by genre and by room, reflecting the comprehensive outcome of the MapReduce operation performed in our wireless sensor network (WSN). This structured approach allows for efficient data retrieval and analysis, significantly enhancing our system's operational capabilities.

*4.6. Middleware Architecture Proposed*

Our middleware architecture is structured into several layers, each designed to address specific aspects of system functionality and operational efficiency:

1. Environment Layer: This layer oversees the system's capabilities and properties, ensuring optimal performance and resource utilization.
   a. Physical Layer: Comprises the actual mini-computers, such as Raspberry Pis, utilized for data storage, inter-device connectivity, and interaction with the remote database. These computers form the hardware foundation of our middleware.
2. Environment Management Layer: Concentrates on the reliability and continuous operation of the system:
   a. Network Layer: Encompasses cloud computing functionalities and communication processes, including establishing connections, performing CRUD (Create, Read, Update, Delete) operations with the database, and managing server–client IP addresses and port information. Communication is facilitated through UDP (User Datagram Protocol) via a WebSocket protocol, ensuring efficient message exchange.
3. Task Management Layer**:** Targets the execution of tasks and management of system dependencies:
   a. Application Layer: Contains the core implementation of our middleware, encompassing algorithm services, endpoints for interaction and connectivity, and the execution of the MapReduce method in both visitor and room count modes, along with the analysis of the resulting data.
   b. Data Layer: Responsible for the local storage of sensor measurements and the aggregation of server results in the event of system interruptions or errors. This layer plays a crucial role in performing ETL (Extract, Transform, Load) operations, ensuring data integrity and availability.

This layered approach to middleware design allows for a clear separation of concerns, making the system more manageable, scalable, and resilient to failures. Each layer serves a distinct purpose, from managing physical resources and network communications to executing specific tasks and handling data storage and analysis, collectively ensuring the smooth operation of the middleware and the effective processing of data.

**5. Results**

Our system's input data are collected from a wireless sensor network (WSN) situated in a building preserved as a museum and cultural activity center. Utilizing the MapReduce paradigm, the system performs analyses based on two types of data pairs: visitor identity (visitor, value) and location (room, value). This dual approach is applied to analyze visitor data and activity within the first four rooms as depicted in the layout presented in Figure 2, excluding Room 5 due to its use for special exhibitions and events.

The system executes the MapReduce operation for monitoring both visitors and rooms, categorizing visitors by their RFID tag (man, woman, or other for inclusivity towards individuals with disabilities and their guide dogs), and tallying the total number of visitors per room. This methodological approach aims to glean insights into visitor demographics and room popularity, aiding in the optimization of museum operations and visitor experience.

The results from these analyses are methodically documented in Table 1 for visitor monitoring and Table 2 for room monitoring. These tables provide a detailed breakdown of visitor flow and room engagement, offering valuable data to museum management for strategic planning and operational adjustments.

**Table 1.** MapReduce algorithm operation for client data sent and server's response for visitor monitoring (man, woman, other).

| |
|---|
| $Client >> man = 1, man = 3, man = 4, man = 2, woman = 3, other = 3, other = 4, other = 3, other = 2, woman = 1, woman = 4, woman = 2, woman = 2, woman = 3, woman = 2, woman = 4 |
| $Server >> {<man, 10>}, {<woman, 21>}, {<other, 12>} |

**Table 2.** MapReduce algorithm operation for client data sent and server's response for room monitoring.

| |
|---|
| $Client >> man = 1, man = 3, man = 4, man = 2, woman = 3, other = 3, other = 4, other = 3, other = 2, woman= 1, woman = 4, woman = 2, woman = 2, woman = 3, woman = 2, woman = 4 |
| $Server >> {<Room1, 2>}, {< Room2, 5>}, {< Room3, 4>}, {< Room4, 5>} |

Table 3 details the performance metrics for low-cost, low-power computing devices, such as Raspberry Pis, within the application layer of our system, focusing on a setup involving five rooms. Each room is equipped with one mini-computing device that serves dual roles as both server and client. We highlight that latency in wireless sensor networks (WSNs) varies, but our data aim to provide an average operational latency metric. It is important to note that latencies exceeding 100–150 ms could affect our system's responsiveness and efficiency during network layer activities, with lower latency values being preferable. Additionally, our measurements indicate that the time to the first byte falls within recommended ranges for software applications, deemed acceptable if below 0.8 s, with an average of 200–600 ms being typical. These metrics are crucial for assessing energy efficiency and performance, playing a key role in evaluating our project's requirements, potential costs, and scalability options, especially under real-life constraints.

**Table 3.** Server–client hardware metrics for low-cost and low-power computing devices (Raspberry Pi model).

| | **Server** | **Client** |
|---|---|---|
| RAM [MB] | 235 | 195 |
| CPU [%] | 64 | 46 |
| Power [A] | 0.52 | |
| Network Time [ms] | Latency = 64 | TTFB = 520.29 |

Similarly, the network layer of our application, which underpins the entire rationale of our algorithm, was evaluated using these mini-computing devices. We measured the average response times for server and client nodes during various stages of operation and documented the number of UDP messages exchanged between the server and client during each phase. These data were facilitated through socket connections, while sensor measurements were captured and stored both locally and remotely. Our findings on system response times and communication metrics, alongside hardware performance measurements under different system loads (triggered by data generation from multiple visitors), are illustrated in Figures 5 and 6.

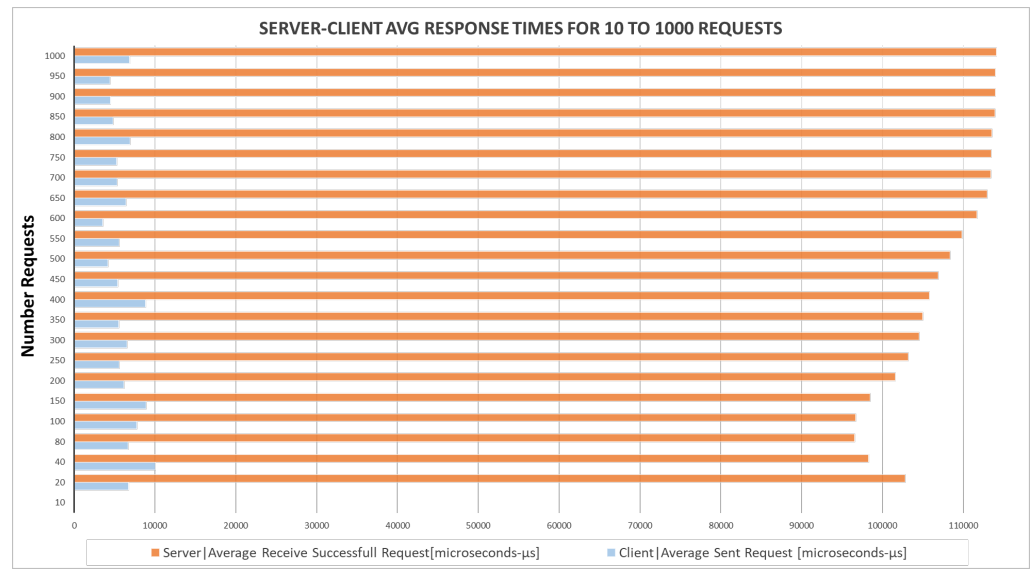

**Figure 5.** Average time (in msec) for a server–client response for 0 to 1500 requests.

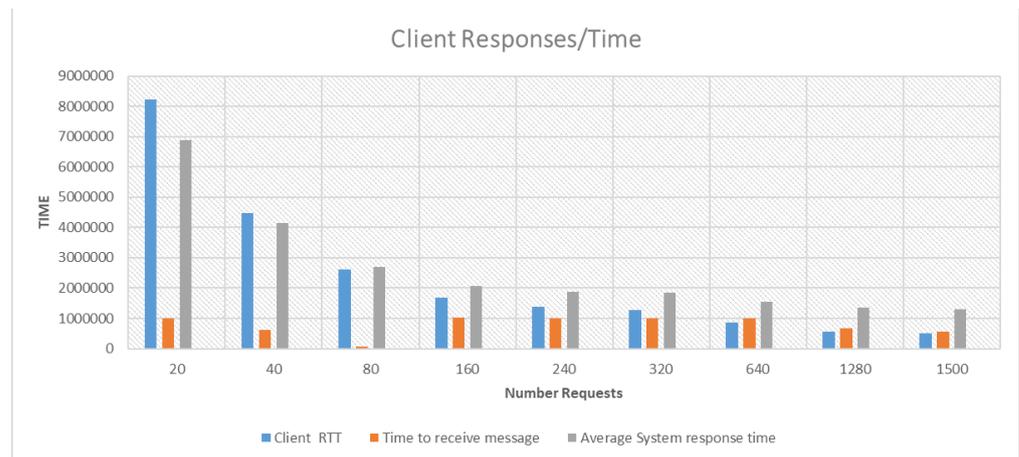

**Figure 6.** Average system response time, round trip time, and time to receive a message for a server–client response for 0 to 1500 requests.

In addition, to assess and understand the execution time and the actual implementation of our application, we present real values for a full software cycle. This includes the initial time for the system to boot up and complete the first five steps of the described algorithm (Figure 3). To achieve accurate performance analysis, we consistently used SD cards with a read/write speed of 80–100/60 MB per second. Additionally, all devices were connected to USB 2.0 ports and used the same drivers. It is noted that using an SSD or a faster SD card with SSD-like speeds (420 MB/s read and 380 MB/s write) would likely reduce the execution times and computer resource consumption significantly. However, our goal is to provide a low-power and, most importantly, low-cost solution suitable for rapid prototyping in the early stages of application development. Therefore, we have not extended our test to SSD storage units.

Specifically, to extract the cycle's metric values and provide a detailed comparison, we executed the application in a headless Java environment using the JVM, transforming Java code to bytecode to optimize machine code execution during runtime [89,90]. Similarly, we optimized the JIT compiler by method inlining (replacing method invocations to avoid the overhead of method calls), escape analysis, and dead code elimination [90,91]. This was carried out to remove unnecessary objects and allocate them on the stack instead of the heap, reducing memory allocation overhead. We also used subexpression elimination to avoid redundant computations. Stack allocation is less resource-demanding than heap allocation as it reduces garbage collection overhead and allows for optimal access to objects [92]. We aimed to create objects locally, avoid passing them to other methods, and keep their scope local when interacting with the database.

Finally, it is worth noting that we excluded certain JVM and JIT profiling tools from our test scenarios. For the JVM, we used Java Mission Control, JVisualVM, and JConsole, as well as JDK basic profiling tools such as Runtime.freeMemory, Runtime.totalMemory, and HPROF (Xrunhprof). For JIT, we ran the application with JIT disabled, monitored CPU usage and RAM consumption, and then compared the results with the JIT-enabled run, ensuring the same input data and test conditions for each Raspberry Pi device. The results of our tests are presented in Table 4 and Figure 7. These results can be compared with several studies [93,94] that use WSNs in indoor buildings. However, their solutions focus mainly on AI or mobile phone-sensing middleware. As such, to assess our results with indoor building projects, it is worth noting the following:

- The placement of sensory devices and the study of the blueprints for placing these devices are similar to the methodology used in monitoring the structure of heritage buildings [95]. Additionally, we studied the implementation of a WSN for pet location monitoring, which provided valuable insights into how to track visitors [96]. Furthermore, we examined the application layer for smart environments on a service middleware for users of public and mediated spaces as a means to track crowds and monitor their activities [97].
- Interesting research that aligns with our perspective on how to count and inspect the information loaded on a system is presented in [98]. Although this research specializes in using the Kafka and Redis frameworks, we have managed to achieve similar behavior for fewer than 1000 users (clients in our case). The results presented in their study are superior, but they leverage three web services and use JSON to process the messages/responses of each client. As such, our approach could be used for testing or validating the ground truth of an application. Similarly, our research can be compared with edge-based monitoring and can be used to achieve similar processes to the one presented for edge-based crowd monitoring using Wi-Fi Beacons [99]. Regarding detailed metrics from other research projects that also use the same mini-device, we achieved similar network delay to [100], which specializes in public transport systems using low-cost IoT devices, and similar results to [101], which proposed a crowd density system.
- We managed to approach the network time delay of [98,102], without using the DBSCAN method or clustering/AI techniques and using less-capable hardware, mainly mini-computers, through the flow of our application. While our approach does not match their optimal workflow, we assert that in the early usage and development stages, their application ground truth validation can use our middleware approach to achieve similar results. This allows for testing with minimal effort and on a tight budget.

**Table 4.** Comparison results assessing server–client measurements of completed software cycle for different visitor counts.

| Visitors | CPU Usage [%] | RAM Consumption [MB] | Power Consumption [A] |
|---|---|---|---|
| 50 | 43.8 | 846 | 0.80 |
| 100 | 44.4 | 850 | 0.80 |
| 300 | 54.4 | 870 | 0.81 |
| 500 | 66.8 | 900 | 0.83 |
| 1000 | 74.6 | 999 | 0.84 |

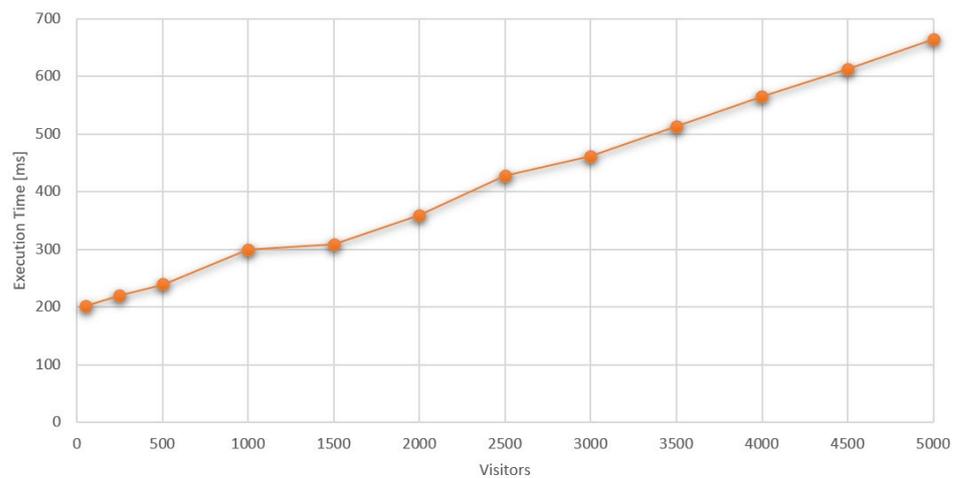

**Figure 7.** Execution time for MapReduce algorithm for room monitoring of different visitor counts.

## 6. Conclusions

MapReduce is widely supported across various programming languages (Python, Java, C#) and offers significant benefits for big-data processing and cloud computing. However, despite its advantages, MapReduce faces limitations, including high latency and limited support for iterative algorithms. Traditionally, MapReduce employs two fault tolerance strategies: controlling and reinitializing the map reducing functions upon failure, and applying checksums to data files to mitigate file corruption. Our innovative approach proposes an alternative method that bypasses checkpointing, advocating for a parallel operational mode in distributed systems. This method is not intended to replace existing mechanisms but to extend MapReduce's applicability, especially in sensor networks with unique characteristics, by focusing on less resource-intensive operations or enhancing node connectivity rather than relying solely on checkpointing for the last successful operation.

Our application of MapReduce is crucial for pinpointing areas of highest interest within a venue, thus facilitating strategic exhibition placement and, in emergency scenarios, tracing virus transmission paths—akin to existing COVID-19 tracking implementations that monitor visitor numbers within a building. The middleware introduced herein utilizes low-cost, low-power computers and demonstrates, through benchmarks, reliable cloud implementation and operation even under hardware constraints.

The technical novelty of our work and the underlying logic of our algorithm empower each mini-computer and sensor node to function interchangeably as either a client or a server. This aspect, coupled with MapReduce implementation, allows for the cost-effective and reliable fusion of data from multiple endpoints in an IoT network. Unlike the majority of case studies focusing on AI or audiovisual analysis, which require expensive resources, our approach facilitates rapid prototyping and swift, scalable solutions across numerous endpoints. Additionally, we propose a fairness index for connectivity status to prevent a single point of failure, automatically monitoring availability and downtime, akin to mainframe computing transactions.

One of the most promising areas for the future of wireless sensor networks (WSNs) is TinyML, which involves ultra-low-power machine learning operations at the edge. Our solution, capable of expanding a cluster of computers to function as both clients and servers or AI operation centers, enables on-device sensor data analytics with minimal power consumption, suitable even for battery-operated devices. Furthermore, our leader election process could be refined to consider additional factors, such as available voltage from a power source or battery/power bank, paving the way for a more sophisticated system that not only manages data (via MapReduce operations) but also other tasks like visualization and system monitoring. This holistic approach aims to leverage each node's capabilities for data distribution and task execution, marking a step towards more complex, efficient WSN management and application.

**Author Contributions:** Both authors made significant contributions to this work. A.G. and E.K. conceived and designed the experiments. A.G. coded the application; was responsible for the investigation, methodology, software, validation, visualization, writing the original draft, and reviewing and editing resources; and carried out the simulation and the optimization and writing—original draft. E.K. contributed to the conceptualization, investigation, methodology, project administration, resources, software, supervision, validation, visualization, review, and editing. All authors have read and agreed to the published version of the manuscript.

**Funding:** This research received no external funding.

**Institutional Review Board Statement:** Not applicable.

**Informed Consent Statement:** Not applicable.

**Data Availability Statement:** The data presented in this study are available on request from the corresponding author.

**Acknowledgments:** This article details our recent work, as part of an MSc course on Distributed and Parallel Systems, at the School of Engineering Department of Democritus University of Thrace, Greece, to apply parallel programming to WSNs. Lastly, the authors are grateful to the civil engineer E. Chamalidou for her assistance in the analysis and the presentation of the floor plans of M. Hatzidakis' residence, as presented in this publication.

**Conflicts of Interest:** The authors declare no conflicts of interest.